\documentclass[showpacs,twocolumn,superscriptaddress]{revtex4}
\usepackage[ansinew]{inputenc}
\usepackage[dvips]{graphicx}
\usepackage{amsmath}
\usepackage{amsthm}
\usepackage{layout}
\usepackage{float}
\usepackage{subfigure}
\usepackage{amsfonts}
\usepackage{amssymb}
\usepackage{bm}
\usepackage{latexsym}
\usepackage{epsfig}
\usepackage{psfrag}

\usepackage{verbatim}

\makeatletter
\def\ExtendSymbol#1#2#3#4#5{\ext@arrow 0099{\arrowfill@#1#2#3}{#4}{#5}}
\def\RightExtendSymbol#1#2#3#4#5{\ext@arrow 0359{\arrowfill@#1#2#3}{#4}{#5}}
\def\LeftExtendSymbol#1#2#3#4#5{\ext@arrow 6095{\arrowfill@#1#2#3}{#4}{#5}}
\makeatother

\begin{document}

\title{Testing Local Realism in $P \to VV$ Decays}

\author{LI JunLi}

%\email{jlli04@mails.gucas.ac.cn}

\affiliation{Dept. of Physics, Graduate University, the Chinese
Academy of Sciences \\ YuQuan Road 19A, 100049, Beijing, China}

\author{QIAO Cong-Feng}

\email{qiaocf@gucas.ac.cn}

\affiliation{Dept. of Physics, Graduate University, the Chinese
Academy of Sciences \\ YuQuan Road 19A, 100049, Beijing, China}

\affiliation{Theoretical Physics Center for Science Facilities
(TPCSF), CAS }

\begin{abstract}
{\bf It was found that the vector meson pair from the pseudoscalar
decays can form an entangled state. In this work we give out
detailed explanations on the polarization correlation of the two
entangled vector mesons. It is demonstrated that an experimental
test of the Clauser-Horne inequality can be carried out through
measuring the azimuthal distribution of four pseudoscalars in the
cascade decay $\eta_c \to VV \to (PP)(PP)$, and the measurement of
this process is feasible with the current running experiments in
tau-charm factory. Moreover, a brief discussion on the polarization
correlation of the two vector mesons from $B \to VV$ decays is also
presented.}
\end{abstract}

\pacs{ 03.65.Ud, 03.67.Mn, 13.25.Ft.}

\maketitle

\section{Introduction}

Although quantum mechanics (QM) represents one of the pillars of
modern physics, the philosophic and physical debates on this
fundamental theory continues ever since its establishment. In the
seminal work \cite{EPR}, Einstein, Podolsky, and Rosen (EPR)
demonstrated that the QM can not provide a complete description of
the ``physical reality" for a two spatially separated but quantum
mechanically correlated particle state which is now known as
entangled state. The premises that were adopted in the EPR's
reasoning can be stated as local realism (LR), where `local' means
the non-existence of action at a distance, and `realism' means that
{\it if, without in any way disturbing a system, we can predict with
certainty (i.e., with probability equal to unity) the value of a
physical quantity, then there exists an element of physical reality
corresponding to this physical quantity.} To avoid the EPR paradox,
it might be a reasonable choice to postulate some additional `hidden
variables' which will restore the completeness and causality to the
theory. This is called the local hidden variable theory (LHVT) that
meets both of the premises of EPR (i.e., LR).

Since it was assumed that the LHVT and QM will lead to the same
observable phenomenology, in the subsequent 30 years, debates
triggered by EPR stay mainly as a matter of philosophical attitude
towards QM. However in 1964, J.S. Bell \cite{Bell1964} showed that
there exist a set of Bell inequalities (BI) which are the
constraints imposed by LHVT and the corresponding QM predictions may
violate these inequalities in some region of parameter space. From
that time on, various forms of Bell's inequalities
\cite{CHSH,CH-inequality} have provided the tool for an experimental
discrimination between QM and LHVT. Many experiments have been
performed mainly using the entangled photon pairs \cite{aspect1,
aspect2, pdc1, pdc2}. All these experiments are substantially in
consistent with the predictions of the standard QM though none of
them can be regard as loophole free \cite{santos-a, santos-la}.
Aiming to get a more conclusive result, and explore the entanglement
with other fundamental interactions \cite{abel-lb} than
electromagnetism, there is an ongoing effort to carry out the
experiment of testing Bell inequality with various physical systems
\cite{ions, B-exp}.

The early attempts of testing LHVT with particle physics concerns
mainly with two 2-dimensional Hilbert space particles. The EPR-like
features of the $K^0\bar{K}^0$ decayed from $J^{\mathrm{PC}} =
1^{--}$ vector mesons had already been noticed in 1960s
\cite{lipkin}. In this case, $K^0\bar{K}^0$ can been considered as
$SU(2)$ doublet, which is called quasi-spins. The entangled state
formed by $K^0\bar{K}^0$ and many other similar neutral meson
systems have been studied since then \cite{bertlmann-rev, KK-06}.
Interesting roles of neutral kaons played in quantum information
theory was studied in \cite{Yu-Shi-1,Shi-Wu}. An experimental test
of Clauser-Horne-Shimony-Holt (CHSH) inequality \cite{CHSH} with
$B^0\bar{B}^0$ pair has been carried out in the $B$ factory
\cite{B-exp}. Based on the data sample of $80\times 10^6$
$\Upsilon(4S) \to B^0\bar{B}^0$ decays at Belle detector at the KEKB
asymmetric collider in Japan, a violation of Bell inequality was
observed, though debate on whether it was genuine test of LHVTs or
not still going on.

On the other hand, T\"ornquist \cite{tor-foud} suggests using the
reaction $e^+e^- \to \Lambda \bar{\Lambda} \to \pi^- p \pi^+
\bar{p}$ to test the quantum correlations of the polarizations
between the baryon pair $\Lambda\bar{\Lambda}$. Similar process
$e^+e^- \to \tau^+ \tau^- \to \pi^+ \bar{\nu}_{\tau}\pi^-
\nu_{\tau}$ was suggested in \cite{abel-lb, ee-tau}. Two typical
processes $\eta_c \to \Lambda \bar{\Lambda} $, $J/\Psi \to \Lambda
\bar{\Lambda}$ were considered in \cite{tor-foud}. Taking $\eta_c$
as an example, the decay distribution of the two pions from
$\Lambda$ decay reads
\begin{eqnarray}
I(\vec{A},\vec{B}) & = & (\frac{ |\rm{S}|^2 + |\rm{P}|^2 }{4\pi})^2
( 1 - \alpha^2 \langle S| \sigma_A \cdot \vec{A} \sigma_B \cdot
\vec{B} |S\rangle ) \nonumber \\ & = &  (\frac{ |\rm{S}|^2 +
|\rm{P}|^2 }{4\pi})^2 ( 1 + \alpha^2 \vec{A} \cdot \vec{B} ) \; ,
\label{tornq-etac}
\end{eqnarray}
where $\vec{A}$ is the unit vector $ \vec{p}_{\pi}^{\text{\,cm}} / |
\vec{p}_{\pi}^{\rm{\,cm}} |$ in the direction of the $\pi^-$
momentum in the $\Lambda$ center of mass frame, $\vec{B}$
corresponds to that of $\pi^{+}$ ; S, P represent the S and P wave
amplitudes; $|S\rangle$ is the spin wave functions of $\eta_c \to
\Lambda \bar{\Lambda}$. T\"ornqvist argued that apart from the
constant $\alpha^2$ and the sign before $\vec{A} \cdot \vec{B}$, the
angular distribution $I(\vec{A},\vec{B})$ represents the correlation
$\langle S| \sigma_A \cdot \vec{A} \sigma_B \cdot \vec{B}
|S\rangle$, and $\vec{A},\vec{B}$ tag the directions of polarization
of $\Lambda,\bar{\Lambda}$. Here the weak decay of $\Lambda \to
\pi^-P (\bar{\Lambda} \to \pi^+\bar{P})$ works as its own
polarimeter. Similar argument exists for $J/\Psi \to
\Lambda\bar{\Lambda}$ and $Z^0 \to \tau^-\tau^+$ cases.

On the experimental side, the DM2 Collaboration \cite{dm2} observed
$7.7 \times 10^6 J/\Psi$ events with about $10^3$ being identified
as from process $J/\Psi \to \Lambda \bar{\Lambda} \to \pi^-P\pi^+
\bar{P}$. Due to the insufficient statistics the experimental
measurement does not give a very significant result. Moreover as
already pointed out by T\"ornqvist, the decay processes, which are
used as the spin analyzer in particle physics, happen spontaneously.
Thus the observer's choice is different from that of the spin
analyzer which can be chosen at will with external polarimeters. A
recent work \cite{Baranov} discussed this issue and expressed the
spin-spin correlations in terms of momentum-momentum correlations
which are experimental measurable quantities, and stated that in the
experimental test, the observer's choice would come in with the
choice of the coordinate system. For more information on the study
of the completeness of QM in high energy physics, readers are
recommended to refer to \cite{dql}. In this work, we plan to give
out more detailed explanation on the measurement of vector meson
entanglement, proposed recently in Ref.\cite{etac-spin-1}.

The structure of the paper goes as follows. In Section 2, using the
method of quantum field theory, we show that the transverse
polarization of the two vectors from $\eta_c$ exclusive decay forms
an entangled state. We focus on our new proposal of testing local
realism with the vector meson pair intermediate state and
demonstrated that this state allows for an experimental test of the
Clauser Horne (CH) inequality \cite{CH-inequality}. In section 3 we
briefly discuss the case of $B$ meson weak decays, i.e., $B \to VV$.
The last section is assigned for summary and conclusions.

\section{Strong decays of $\eta_c \to V V$}

\subsection{The correlation described in quantum theory}

\begin{figure}[t,m,u]
\centering \scalebox{0.4}{\includegraphics{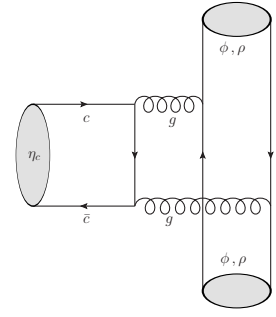}}
\caption{\small The schematic Feynman diagram of processes $\eta_c
\to VV$. Here V stands for vector meson $\phi$, $\rho$, etc.}
\label{feynman-etac}
\end{figure}
\begin{figure}[t,m,u]
\centering \scalebox{0.6}{\includegraphics{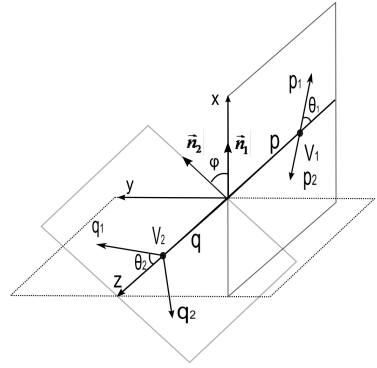}}
\caption{\small Relative angles and decay kinematics in the rest
frame of $V_1$.} \label{angles-etac}
\end{figure}

In quantum field theory, under the constraints of Parity
conservation and Lorentz invariance the decay amplitude of $\eta_c
\to V_1(p,\epsilon^{*}) V_2(q,\epsilon'^*)$, see
Fig.(\ref{feynman-etac},\ref{angles-etac}), takes the following form
\begin{eqnarray}
\mathcal{A}_{\eta_c\to V_1V_2} = i  S \epsilon^{*\mu}
\epsilon'^{*\nu} \varepsilon_{\mu\nu\rho\sigma} \frac{p^{\rho}
q^{\sigma}}{p\cdot q} \;,
\end{eqnarray}
where $S$ is a scalar amplitude and we do not really mind its
details in the aim of entanglement analysis. The above amplitude can
be decomposed of the helicity amplitudes of the final states, like
\begin{eqnarray}
\mathcal{A}_{\pm} = \mathcal{A}(\eta_c \to
V_{1}(p,\epsilon^*_{\pm}) V_{2}(q,\epsilon'^*_{\pm}))= \pm iS\; .
\end{eqnarray}
Here, we choose $q$ to be directed in positive $z$-direction in the
$\eta_c$ rest frame, and the polarization four-vector of the light
vector mesons such that in a frame where both light mesons have
momentum along the $z$-axis, they are
\begin{eqnarray}
\epsilon'^{\mu}_{\pm} = \epsilon^{\mu}_{\mp} = \frac{1}{\sqrt{2}}(
0,\pm 1, i,0 ) \; .
\end{eqnarray}

The decay amplitude of $\eta_c \to
V_1(p,\epsilon^*)V_2(q,\epsilon'^*) \to [P(p_1)P(p_2)],
[P(q_1)P(q_2)]$ can now be expressed as
\begin{eqnarray}
\frac{\mathrm{d}\Gamma_{\eta_c \rightarrow V_1V_2\rightarrow
\ldots}}{\mathrm{d} \cos\theta_1 \mathrm{d} \cos\theta_2
\mathrm{d}\varphi} & \propto & | \sum_{\lambda = \pm 1} A_{\lambda}
\epsilon_{\lambda}^{\mu} \epsilon_{\lambda}'^{\nu} p_{1\mu} q_{1\nu}
|^2 \nonumber \\ & = & | (A_{+} \epsilon_{+}^{\mu}
\epsilon_{+}'^{\nu} + A_{-}\epsilon_{-}^{\mu} \epsilon_{-}'^{\nu})
p_{1\mu} q_{1\nu} |^2 \nonumber \\ & \propto & | (
\epsilon_{+}^{\mu} \epsilon_{+}'^{\nu} - \epsilon_{-}^{\mu}
\epsilon_{-}'^{\nu}) p_{1\mu} q_{1\nu} |^2 \nonumber
\\ & \propto &  | \langle \vec{r}_{p_{1}}|\langle
\vec{r}_{q_1}|\Psi\rangle |^2 \; , \label{D-decay-am}
\end{eqnarray}
with
\begin{eqnarray}
|\Psi\rangle & = & \frac{1}{\sqrt{2}}(|\epsilon_1^\mu\rangle
|{\epsilon'}_2^{\nu}\rangle - |\epsilon_2^\mu\rangle
|{\epsilon'}_1^{\nu}\rangle) \label{polarization-entangled} \; , \\
\vec{r}_{p_{1}} & = & (\sin\theta_1, 0, -\cos\theta_1) =
(\sin\theta_1 \vec{n}_1, -\cos\theta_1) \; ,
\\ \vec{r}_{q_{1}} & = & (\sin\theta_2\cos\varphi, \sin\theta_2\sin\varphi,
\cos\theta_2) \nonumber \\ & = & ( \sin\theta_2\vec{n}_2 ,
\cos\theta_2)\; .
\end{eqnarray}
Here, $\epsilon_1^{\mu} = (0,1,0,0) \; , \; \epsilon_2^{\nu} =
(0,0,1,0)$; $\vec{n}_{1,2}$ are unit vectors in the $x$-$y$ plane
projected by $\vec{r}_{1,2}$; $\theta_{1,2}$ both range from $0$ to
$\pi$; $\varphi$ ranges from $0$ to $2\pi$, as shown in
Fig.(\ref{angles-etac}). Obviously, the wave function $|\Psi
\rangle$ composed of two transverse polarization vectors is an
entangled state.

Integrating over $\theta_{1,2}$ we have
\begin{eqnarray}
\frac{\mathrm{d}\Gamma_{\eta_c \rightarrow V_1V_2\rightarrow
\ldots}}{\mathrm{d}\varphi} & \propto & | \langle
\vec{n}_{1}|\langle \vec{n}_{2}|\Psi\rangle |^2 \; , \label{prob-decay-width} \\
P(\vec{n}_1, \vec{n}_2) & \equiv & | \langle \vec{n}_{1}|\langle
\vec{n}_{2}|\Psi\rangle |^2  \; , \label{qm-def-p}
\end{eqnarray}
where (\ref{qm-def-p}) is the QM definition of the probability of
one particle polarized in direction $\vec{n}_1$ and the other in
direction $\vec{n}_2$.

\subsection{The test of Bell inequality}

We have got the QM prediction for the probability $P(\vec{n}_1,
\vec{n}_2)$ (i.e., Eq.(\ref{qm-def-p})). In the following we show
that these predictions violate the Bell inequality imposed by LHVTs.
To proceed the analysis, we first reformulate the entangled state of
(\ref{polarization-entangled}) in a more compact form. Further
observations of Eq.(\ref{polarization-entangled}) indicate that it
describes the wave function similar to that of entangled photon
pairs in \cite{Bohm}.

Because of the rotation invariance about $z$-axis, we can write
Eq.(\ref{polarization-entangled}) in the following form
\begin{eqnarray}
|\Psi\rangle = \frac{1}{\sqrt{2}}(\, |\epsilon(\varphi)\rangle_{V_1}
|\epsilon(\varphi_{\perp})\rangle_{V_2} -
|\epsilon(\varphi_{\perp})\rangle_{V_1}
|\epsilon(\varphi)\rangle_{V_2}) \label{epsilon-state} \; ,
\end{eqnarray}
where $\epsilon(\varphi)$ is polarization vector in direction
$\varphi$ (see Fig.(\ref{angles-etac})), and $\varphi_{\perp} =
\varphi+\frac{\pi}{2}$. Since the transverse polarization of vector
meson has two degrees of freedom, from (\ref{epsilon-state}) we can
infer that if $V_1$ is polarized along the direction $\varphi$, the
polarization of $V_2$ is then determined simultaneously: it must be
polarized perpendicular to that of $V_1$.

Suppose that the transverse polarization of the state
(\ref{epsilon-state}) is completely specified by a set of parameters
$\lambda$, and the probabilities of a count being triggered by the
decays of $V_{1}, V_{2}$ polarizing along $\vec{n}_{1}$ and
$\vec{n}_{2}$ are $p(\vec{n}_1,\lambda)$ and $q(\vec{n}_2,\lambda)$,
respectively. According to LR, the joint probability of particle
$V_1$ polarizing along $\vec{n}_1$ and particle $V_2$ in $\vec{n}_2$
is given by
\begin{eqnarray}
P(\vec{n}_1,\vec{n}_2) = \int p(\vec{n}_1, \lambda) q(\vec{n}_2,
\lambda) \rho(\lambda) \, \mathrm{d} \lambda \; ,
\end{eqnarray}
and the single side probabilities are
\begin{eqnarray}
P(\vec{n}_{1}) = \int p(\vec{n}_1, \lambda) \rho(\lambda) \,
\mathrm{d} \lambda \; , \label{def-Pn1}\\ P(\vec{n}_{2}) = \int
q(\vec{n}_2, \lambda) \rho(\lambda) \, \mathrm{d} \lambda \; ,
\label{def-Pn2}
\end{eqnarray}
where $\int \rho(\lambda)\, \mathrm{d} \lambda = 1$. Using the
simple algebraic theorem \cite{CH-inequality}
\begin{eqnarray}
-XY \leq xy -xy' + x'y + x'y' -x'Y - Xy \leq 0 \; ,
\end{eqnarray}
where $x,x',y,y',X,Y$ are real numbers, and $0\leq x,x' \leq X,
0\leq y,y' \leq Y $, and substituting $x,y$ with
$p(\vec{n}_1,\lambda), q(\vec{n}_2,\lambda)$, setting $X,Y=1$, one
can readily get the CH inequality
\begin{eqnarray}
P(\vec{n}_{1},\vec{n}_{2}) - P(\vec{n}_{1},\vec{n}'_{2}) +
P(\vec{n}'_{1}, \vec{n}_{2}) & & \nonumber \\  + P(\vec{n}'_{1},
\vec{n}'_{2}) - P(\vec{n}'_1) - P(\vec{n}_2) & \leq & 0 \; .
\label{CH-v2}
\end{eqnarray}
This stands as a  constraint on $P(\vec{n}_1,\vec{n}_2)$s imposed by
LR. Substituting the quantum mechanical predictions (\ref{qm-def-p})
into the inequality (\ref{CH-v2}), we arrive at
\begin{eqnarray}
\frac{1}{2} [\, \sin^2(\varphi_{1} - \varphi_{2} ) -
\sin^2(\varphi_{1} - \varphi'_{2} ) & & \nonumber \\  +
\sin^2(\varphi'_{1} - \varphi_{2} ) + \sin^2(\varphi'_{1} -
\varphi'_{2} ) - 2 \, ] & \leq & 0\; . \label{ch-qm-result}
\end{eqnarray}
Here, $\varphi\,$s are the azimuthal angles of the $\vec{n}\,$s.
(\ref{ch-qm-result}) is easily found to be violated while
\begin{eqnarray}
\varphi'_{1} = \varphi_{1} + \frac{\pi}{4} \; , \varphi_{2} =
\varphi_{1} + \frac{5\pi}{8} \; , \varphi'_{2} = \varphi_{1} +
\frac{7\pi}{8} \; , \label{angles38}
\end{eqnarray}
which gives $\frac{\sqrt{2}-1}{2} \leq 0$.

The inequality (\ref{CH-v2}) bears no additional assumptions.
However, if one adopts the {\it no-enhancement} assumption
\cite{CH-inequality}, the CH inequality then takes the following
form
\begin{eqnarray}
-P(\infty, \infty) & \leq & P(\vec{n}_{1}, \vec{n}_{2}) -
P(\vec{n}_{1}, \vec{n}'_{2})\nonumber
\\ & & + P(\vec{n}'_{1}, \vec{n}_{2}) + P(\vec{n}'_{1}, \vec{n}'_{2}) \nonumber \\ & & - P(\vec{n}'_{1},
\infty) - P(\infty, \vec{n}_{2}) \leq 0 \label{no-en-ch} \; ,
\end{eqnarray}
where symbol $\infty$ denotes the absence of analyzer on the
corresponding side. If we further assume that $P(\vec{n}_1,\infty) =
P(\vec{n}_1,\vec{n}_{2}) + P(\vec{n}_1,\vec{n}_{2\perp})$
\cite{torgerson, EI-Garuccio}, then (\ref{no-en-ch}) gives
\begin{eqnarray}
P(\vec{n}'_{1}, \vec{n}_{2}) \leq P(\vec{n}_{1\perp} , \vec{n}_{2})
+ P(\vec{n}'_{1}, \vec{n}'_{2\perp}) + P(\vec{n}_{1}, \vec{n}'_{2})
\label{ineq-non-en} \; .
\end{eqnarray}
Here, $\vec{n}_{1\perp}, \vec{n}'_{2\perp}$ are the orthogonal
directions to $\vec{n}_1, \vec{n}'_{2}$, and their azimuthal angles
satisfy
\begin{eqnarray}
\varphi_{1\perp} = \varphi_{1} + \frac{\pi}{2} \; ; \;
\varphi'_{2\perp} = \varphi'_{2} + \frac{\pi}{2} \; .
\end{eqnarray}
Similar as (\ref{ch-qm-result}), inputting the quantum mechanical
results into inequality (\ref{ineq-non-en}), we have
\begin{eqnarray}
\frac{1}{2}\{ \sin^2( \varphi'_{1} - \varphi_{2} ) - [ \cos^{2}(
\varphi_{1} - \varphi_{2} ) & &\nonumber \\  + \cos^2(\varphi'_{1} -
\varphi'_{2}) + \sin^{2}(\varphi_{1} - \varphi'_{2}) ] \} & \leq & 0
\label{viol-non-en} \; ,
\end{eqnarray}
which is numerically equivalent to inequality (\ref{ch-qm-result}),
and it is also violated by quantum mechanics while
\begin{eqnarray}
\varphi'_{1} = \varphi_{1} + \frac{\pi}{4} \; , \varphi_{2} =
\varphi_{1} + \frac{5\pi}{8} \; , \varphi'_{2} = \varphi_{1} +
\frac{7\pi}{8} \; .
\end{eqnarray}

It is noteworthy that there are some differences between inequality
(\ref{CH-v2}) and (\ref{ineq-non-en}). The (\ref{CH-v2}) is an
inhomogeneous one which contains both coincidence and single
probabilities, while (\ref{ineq-non-en}) is a homogeneous one which
is merely composed of several coincidence probabilities
\cite{santos-a}. From a practical point of view, the homogeneous
inequality allows test involving only coincidence counting rates.
The inequality will be insensitive to many scale factors like
detector efficiencies in this case and is extremely convenient for
practical experiment. However the derivation of homogeneous
inequalities requires additional assumptions besides locality and
realism \cite{santos-a, homoge}, for more discussions of this issue
we refer to a recent work \cite{homoge}.

In the experiment, given that the four final pseudoscalars move with
momenta $p_1, p_2, q_1, q_2$, the azimuthal angle $\varphi$ between
two decay planes of the entangled vector meson pair equals to the
angle between $\vec{n}_1$ and $\vec{n}_2$, as shown in
Fig.(\ref{angles-etac}). The magnitudes of $P(\vec{n}_1, \vec{n}_2)$
in the CH inequality are therefore experimentally measurable, which
is obviously the probability density, up to an overall normalization
factor $\xi$, from the definition of $P(\vec{n}_1, \vec{n}_2)$. That
is
\begin{eqnarray}
P(\vec{n}_1, \vec{n}_2)/\xi_{p} = N'(\varphi)/(\xi_{n} \cdot N) \; .
\label{prop-P-N}
\end{eqnarray}
Here, $\xi_{p}$ and $\xi_{n}$ satisfy the following normalization
condition
\begin{eqnarray}
\int_{0}^{2\pi} \int_{0}^{2\pi} P(\vec{n}_1, \vec{n}_2) \,
\frac{\mathrm{d} \vec{n}_1}{\sqrt{\xi_{p}}}
\frac{\mathrm{d}\vec{n}_2}{\sqrt{\xi_{p}}} & = & 1 \; ,
\\ \int_{0}^{2\pi} \int_{0}^{2\pi} N'(\varphi)/N \, \frac{\mathrm{d}
\vec{n}_1}{\sqrt{\xi_{n}}} \frac{\mathrm{d}
\vec{n}_2}{\sqrt{\xi_{n}}} & = & 1\; ,
\end{eqnarray}
with
\begin{eqnarray}
N'(\varphi) & = & \frac{N(\varphi+\Delta \varphi) -
N(\varphi)}{\Delta \varphi} \; ,
\end{eqnarray}
where $N(\varphi)$ is the event number within azimuthal angle
$\varphi$ -- the angle between $\vec{n}_1$ and $\vec{n}_2$, $N$ is
the total event number. Eq.(\ref{prop-P-N}) can be expressed in a
more simpler form
\begin{eqnarray}
P(\vec{n}_1, \vec{n}_2) & = & \kappa \cdot N'(\varphi)/N \; .
\label{exp-pnn}
\end{eqnarray}
where $\kappa = \xi_{p}/\xi_{n}$. It can be easily obtained that
$\kappa = \frac{\pi}{2}$, because in computing an isolated
probability $P(\vec{n}_1, \vec{n}_2)$, the LR models should give the
same results as quantum mechanics. And, it can be seen from
Eq.(\ref{epsilon-state}) that there will be two possible outcomes
($\varphi,\varphi_{\perp}$) if a polarization analyzed decay process
happened, thus the single side probability $P(\vec{n}_{1,2})$ can be
measured through
\begin{eqnarray}
P(\vec{n}_{1,2}) = \frac{N'(\varphi_{1,2})}{N'(\varphi_{1,2}) +
N'(\varphi_{1,2}+\pi/2)} \; . \label{exp-pn}
\end{eqnarray}

In the above expressions, apart from the constant $\kappa$, the
right hand sides of Eqs. (\ref{exp-pnn}) and (\ref{exp-pn}) are
experimentally measurable, i.e., $N'(\varphi)/N$ is the differential
decay width of $\eta_{c}$ to four pseudoscalar mesons divided by its
total width via intermediate vector mesons. Inputting the
experimental results of (\ref{exp-pnn}) and (\ref{exp-pn}) in the
configuration of (\ref{angles38}) into (\ref{CH-v2}), one may in
principle find the incompatibility of quantum theory with LR.
However, in practice, to perform the test of incompatibility the
experiment efficiency should be taken into account. The general
inequality efficiency and background levels were once discussed by
Eberhard \cite{Eberhard-ineq}, and for the wave function
(\ref{epsilon-state}) the violation of inequality (\ref{CH-v2})
yields the threshold efficiency  $ \eta > 82.8\%$
\cite{Brunner-Simon}.

To carry out the test of Bell type inequality, the decay angles
$\varphi$s should generally be chosen actively by experimenters, but
this is not the case for mesons due to the passive character of
their decays. Thus, here only a restricted class of LR can be tested
\cite{Tsubasa-Izumi}. A genuine Bell test also requires the decay
events of two vector mesons $V_1$ and $V_2$ to be space-like
separated. For the strongly decayed vector mesons ($\phi,\rho$,
etc.), it is difficult to spatially distinguish the vertexes between
them. Thus one can not guarantee for each particular event of
$\eta_c \to VV \to (PP)(PP)$ that the decays of the two vector
mesons are separated space-likely. However, one can obtain the
fraction of space-like separation events over the total events.
Given $x_1$ and $x_2$ the distance (in $\eta_c$ rest frame) from the
$\eta_c$ decay point to the decay points of two $\phi$s (or
$(\rho\rho)$, etc.), the space-like condition is \cite{tor-foud}
\begin{eqnarray}
\frac{1}{k} \leq \frac{x_1}{x_2} \leq k \; ,
\end{eqnarray}
where $k = \frac{1+\beta_V}{1-\beta_V}$, $\beta_V = \frac{v_{V}}{c}
= \sqrt{1-1/\gamma^2}$, $\gamma=E_V/m_V = E_{\eta_c}/2m_V$. The
fraction of the space-like separated decay events to total events of
the vector meson pairs is
\begin{eqnarray}
\text{F}  =  \int_0^{\infty} e^{-x_2} \, \mathrm{d}x_2
\int_{\frac{1}{k}x_2}^{kx_2} e^{-x_1} \, \mathrm{d}x_1 = \frac{k-1}{
k + 1} \; .
\end{eqnarray}
Obviously, the fraction F equals to $\beta_{V}$ \cite{tor-foud}. For
the space-like events, constraint imposed on
$P(\vec{n}_1,\vec{n}_2)$s by the restricted class of local realism
is given by inequality (\ref{CH-v2}), where the upper limit is zero.
As for the non-space-like(time-like) events, the upper limit of the
left hand side of (\ref{CH-v2}) can maximally amount to $1/2$. In
the mixture of space-like and non space-like events with ratios of
$\beta_{V}$ and $1 - \beta_{V}$, the upper limit of the left hand
side of (\ref{CH-v2}) is \cite{Tsubasa-Izumi}
\begin{eqnarray}
0 \cdot \beta_V + \frac{1}{2} \cdot (1 - \beta_V)\; .
\end{eqnarray}
Therefore, to carry out the test of local realism the lower bound
for the ratio $\beta_{V}$ is $\geq 2-\sqrt{2}$.

Specifically, the processes $\eta_c \to \phi\phi$ and $\eta_c \to
\rho\rho$ are well-established, and the subsequent two body decays
of them are
\begin{eqnarray}
\phi \to K^+ K^- (K_L^0 K_S^0) \; ,\; \rho \to \pi\pi \; ,
\label{branching-f}
\end{eqnarray}
with large branching fractions: $\phi\rightarrow K^+ K^-
(K_L^0K^{0}_S)$ $\sim 49.2\pm0.6\% \,(34.0\pm0.5\%)$ and
$\rho\rightarrow \pi\pi$ $\sim 100\%$ \cite{PDG}. The magnitudes of
$\beta_V$ for $\phi(1020)$ and $\rho(770)$ in $\eta_c$ decay are
$\beta_{\phi} = 0.729$ and $\beta_{\rho} = 0.856$, which are both
larger than the lower bound of $2-\sqrt{2} \approx 0.586$.

\section{The polarization correlation emerged in $B\rightarrow V_1V_2$}

Now we turn to the polarization correlation of the vector mesons in
$B$ meson exclusive weak decays, that is $B\to V_1V_2$. There is
special interest in the analysis of this process, because it is
well-known that the parity is violated in the weak interaction.

The full angular dependence of the cascade decay where both vector
mesons decay into pseudoscalar particles is given by \cite{bvv}
\begin{eqnarray}
& & \frac{\mathrm{d}\Gamma_{B\rightarrow V_1V_2\rightarrow
\cdots}}{\mathrm{d} \cos\theta_1 \mathrm{d} \cos\theta_2
\mathrm{d}\varphi} \propto  \nonumber \\ & & |\mathcal{A}_0|^2
\cos^2\theta_1 \cos^2\theta_2 + \frac{1}{4} \sin^2\theta_1
\sin^2\theta_2 ( | \mathcal{A}_+ |^2 + | \mathcal{A}_{-} |^2 )
\nonumber \\ & &  - \frac{1}{4} \sin2\theta_1 \sin2\theta_2 [
\mathrm{Re}(e^{-i\varphi} \mathcal{A}_0 \mathcal{A}^{*}_+) +
\mathrm{Re}(e^{+i\varphi} \mathcal{A}_0 \mathcal{A}^{*}_-) ]
\nonumber \\ & &  + \frac{1}{2} \sin^2\theta_1 \sin^2\theta_2
\mathrm{Re}( e^{+2i\varphi} \mathcal{A}_+ \mathcal{A}^{*}_-)
\label{full-angular-distribution}
\end{eqnarray}
where $\mathcal{A}_{0,\pm}$ are the helicity amplitudes. Five
observables corresponding to three amplitudes and two relative
phases of the helicity amplitudes are well defined. The typical set
of observables consists of the branching fraction, two out of the
three polarization fractions $f_{L}, f_{\parallel}, f_{\perp}$, and
two phases $\phi_{\parallel}, \phi_{\perp}$, where
\begin{eqnarray}
f_{L, \parallel, \perp} = \frac{|\mathcal{A}_{0, \parallel,
\perp}|^2}{ |\mathcal{A}_0|^2 +|\mathcal{A}_{\parallel}|^2 +
|\mathcal{A}_{\perp}|^2 } \; , \; \phi_{\parallel,\perp}^{B} =
\mathrm{arg}\,\frac{\mathcal{A}_{\parallel, \perp}}{\mathcal{A}_0}
\;  \label{b-5-observables}
\end{eqnarray}
with
\begin{eqnarray}
\mathcal{A}_{\parallel}  =  \frac{\mathcal{A}_{+} +
\mathcal{A}_{-}}{\sqrt{2}} \; ; \; \mathcal{A}_{\perp}  =
\frac{\mathcal{A}_{+} - \mathcal{A}_{-}}{\sqrt{2}} \; .
\end{eqnarray}
One can then obtain the azimuthal angle dependence from the
subsequent decays of $V_1$ and $V_2$.

In $B$ decays, integrating over $\theta_1$ and $\theta_2$, the
Eq.(\ref{full-angular-distribution}) becomes
\begin{eqnarray}
& & \frac{\mathrm{d}\Gamma_{B\rightarrow V_1V_2\rightarrow
\ldots}}{\mathrm{d}\varphi}  \propto  \frac{4}{9}( |\mathcal{A}_0|^2
+ | \mathcal{A}_{+}e^{i\varphi} + \mathcal{A}_{-}e^{-i\varphi} |^2 )
\nonumber \\ & = & \frac{4}{9}( |\mathcal{A}_0|^2 + 2|
\mathcal{A}_{\parallel}\cos\varphi + i
\mathcal{A}_{\perp}\sin\varphi |^2 )\; .
\label{integrated-over-theta}
\end{eqnarray}
Substituting (\ref{b-5-observables}) into
(\ref{integrated-over-theta}), we have
\begin{eqnarray}
& & \frac{\mathrm{d}\Gamma_{B\rightarrow V_1V_2\rightarrow
\ldots}}{\mathrm{d}\varphi}   \propto  ( 1 + \cos2\varphi \cdot
(f_{\parallel} - f_{\perp}) \nonumber \\ & & + 2\sin2\varphi
\sin(\phi_{\parallel} - \phi_{\perp}) \sqrt{f_{\parallel} f_{\perp}}
) \; . \label{correlation-in-weak}
\end{eqnarray}
Taking the same procedure as introduced in Section 2.2, we have
\begin{eqnarray}
P(\vec{n}_1,\vec{n}_2) & = & \frac{1}{4}( \frac{ 2\pi \cdot
\mathrm{d} \Gamma_{B\rightarrow V_1V_2\rightarrow \ldots}}{
\mathrm{d} \varphi
\cdot \Gamma_{B \rightarrow V_1 V_2\rightarrow \ldots}} - f_{L}) \nonumber \\
& = & \frac{1}{4}( f_{\parallel} + f_{\perp} + \cos2\varphi \cdot
(f_{\parallel} - f_{\perp}) \nonumber \\ & & + 2 \sin2\varphi \cdot
\sin(\phi_{\parallel} - \phi_{\perp}) \cdot
\sqrt{f_{\parallel}f_{\perp}} ) \label{P-B-vv} \; ,
\end{eqnarray}
where $\varphi$ is the azimuthal angle between $\vec{n}_1$ and
$\vec{n}_2$.

The observables in (\ref{P-B-vv}) are obtainable in the experiment.
For instance, in the experiment the process $B^0 \rightarrow \phi
K^{0*}$ tells \cite{b0-phik}: $f_L = 0.52\pm0.05\pm0.02$, $f_{\perp}
= 0.22\pm0.05\pm0.02$, $f_{\parallel} = 1 - f_L - f_{\perp}=0.26 $,
$ \phi_{\parallel} = 2.34 $, $ \phi_{\perp} = 2.47 $. Hence, the
$P(\vec{n}_1,\vec{n}_2)$ in (\ref{P-B-vv}) reads
\begin{eqnarray}
P(\varphi_1,\varphi_1+\varphi) & = & \frac{1}{4}( 0.48 +
0.04\cos2\varphi \nonumber \\ & & - 2\cdot 0.24 \cdot \sin2\varphi
\sin(0.13) ) \; . \nonumber
\end{eqnarray}
Obviously, the polarization correlation is suppressed in this case
due to the reasons of $f_{\parallel} \approx f_{\perp}$ and $
\phi_{\parallel} \approx \phi_{\perp}$.

\section{Summary and Concluding Remarks}

In this work we have investigated the EPR-like correlations of the
entangled vector meson pair in $\eta_c$ and $B$ decays. Contrary to
the measurement of correlation function of polarization, which were
suggested to perform in the processes of $e^+e^- \to
\Lambda\bar{\Lambda}$ and $e^+e^- \to \tau^-\tau^-$, the CH
inequality for the experimental test proposed in this work involves
only the probabilities of transverse polarization states of the
vector mesons. This reaps the benefit of the boost invariance of
transverse polarization along the momentum direction of any one of
the two vector mesons in $\eta_c$ rest frame. The probabilities in
the CH inequality are shown to be experimentally measurable through
subsequent two-body decays of the vector mesons, $\phi$, $\rho$,
etc. Since the measurements on $\phi$ or $\rho$ to two-pseudoscalar
decays are well established in the experiment, and all these decays
possess large branching fractions, the $P \to VV$ processes
therefore enable us to perform the test of local realism in current
running colliders.

It should be mentioned that the passive character of the particle
decays and the non-space-like decay events may induce restrictions
on the LR models being tested and the so-called locality loophole to
the experiment, which hinder the proposed test to refute the LR
definitly. Nevertheless, the experimental realization of the
proposals in this work may extend the test of nonlocality into the
high energy regime with high dimensions, which will give us a more
explicit conclusion in comparison with that from bipartite qubit
case.

Moreover, taking into account proposals \cite{tor-foud} and
\cite{ee-tau}, the experimental tests of the Bell inequalities
involving spin or polarization in elementary particle physics can
now be assorted into four classes, i.e.
\begin{eqnarray}
\eta_c \stackrel{\text{Strong}}{\longrightarrow} & \Lambda
\bar{\Lambda} & \stackrel{\text{Weak}}{\longrightarrow} \pi^-p \;
\pi^+ \bar{p} \; ,\label{sw}\\ Z^0(\gamma^*)
\stackrel{\text{Weak(QED)}}{\longrightarrow} & \tau^- \tau^+ &
\stackrel{\text{Weak}}{\longrightarrow} \pi^-\nu_{\tau} \; \pi^+
\bar{\nu}_{\tau}\; , \label{ww}\\ \eta_c
\stackrel{\text{Strong}}{\longrightarrow} & \phi \phi &
\stackrel{\text{Strong}}{\longrightarrow} K\bar{K} \; K\bar{K}\; , \label{ss}\\
B \stackrel{\text{Weak}}{\longrightarrow} & \phi K^{*} &
\stackrel{\text{Strong}}{\longrightarrow} K\bar{K} \; K\pi \;
.\label{ws}
\end{eqnarray}
The (\ref{sw}) corresponds to the suggestion of T\"ornqvist
\cite{tor-foud}; (\ref{ww}) corresponds to the proposal of Privitera
{\it et al.} \cite{ee-tau}; (\ref{ss}) and (\ref{ws}) belong to ours
in this work. Each of the above processes in fact undergoes two
steps. The first step can be viewed as the entanglement generation
process, while the second step can be interpreted as the process of
spin analyzing. In those two steps, either strong or weak
interaction plays the dynamical role. The four different
combinations of strong and weak interactions in the two steps are
exhibited in Eqs.(\ref{sw})-(\ref{ws}). Taking into account the
photonic experiment, which is dominated by electromagnetic
interaction, proposals for testing Bell inequalities have been put
forward in three of the four fundamental interactions. To our best
of knowledge the only fundamental interaction which has not been
employed to generate and detect quantum entanglement is gravity.

%%%%%%%%%%%%%%%%%%%%%%%%%%%%%%%%%%%%%%%%%%%%%%%%%%%%%%%%%%%%%%%%%%%%%

\vspace{.2cm} %{\bf Acknowledgments} \vspace{.3cm}

This work was supported in part by the National Natural Science
Foundation of China(NSFC) under the grants 10821063 and 10775179, by
CAS Key Project on "$\tau$-Charm Physics(NO.KJCX2-yw-N29), by the
Scientific Research Fund of GUCAS, and by the Project of Knowledge
Innovation Program (PKIP) of CAS with Grant No. KJCX2.YW.W10.
%%%%%%%%%%%%%%%%%%%%%%%%%%%%%%%%%%%%%%%%%%%%%%%%%%%%%%%%%%%%%%%%%%%%%

\end{document}